# Three Controlled Experiments in Software Engineering with the Two-Tier Programming Toolkit: Final Report

Amnon H. Eden, Epameinondas Gasparis
School of Computer Science & Electronic Engineering, University of Essex

Revision history:

- Minor corrections: 5 Jan. 2010
- Typos corrected: 8 July 2009
- Final report published: 21 May 2009
- First published as an interim report: 16 May 2009

## Synopsis

Three controlled experiments testing the benefits that Java programmers gain from using the Two-Tier Programming Toolkit have recently been concluded. The first experiment offers statistically significant evidence (p-value: 0.02) that programmers who undertook only minimal (1-hour) training in using the current prototype exhibit 76% productivity gains in key tasks in software development and maintenance. The second experiment shows that the use of the TTP Toolkit is likely (p-value: 0.10) to almost triple the accuracy of programmers performing tasks associated with software quality. The third experiment shows that the TTP Toolkit does not offer significant productivity gains in performing very short (under 10 min.) tasks.

Table of Contents

## Acknowledgements

This experiments were funded by a grant from the Research Promotion Fund, School of Computer Science & Electronic Engineering, University of Essex. They were designed, planned, set up, and conducted by Epameinondas. We wish to thank Jonathan Nicholson and Christina Maniati for their help in designing and conducting the experiments, Rick Kazman and Udo Kruschwitz for their advice on methodology.

## General

These controlled experiments were designed to test the productivity gains that can be expected from using the the abstraction, visualization, and design verification mechanisms provided by the Two-Tier Programming technology during software development and maintenance, as well as the technology's contributions to software quality. The technology is demonstrated by version 0.5.3 of the Two-Tier Programming Toolkit (henceforth: the TTP Toolkit) developed by members of the [Two-Tier Programming Project](). (The current prototype is also available for [download]() under a under a Creative Commons Licence). Programmers' performance was measured in carrying out software [comprehension](), [conformance]() ("design verification"), and [evolution]() tasks using the TTP Toolkit compared to their performance in carrying out same tasks using the standard Integrated Development Environment for Java programs, Sun's NetBeans version 6.1 (henceforth: NetBeans). Participants were graduate (MSc or PhD) students at the School of Computer Science & Electronic Engineering, University of Essex. At the beginning of each experiment, participants received one hour training in NetBeans and one hour training in the TTP Toolkit in carrying out the relevant tasks. After concluding their training, participants were divided into a control group (using NetBeans) and an experiment group (using the TTP Toolkit), and all participants carried out same task. To minimize bias, participants switched groups between the tasks; in other words, participants who used the TTP Toolkit (the experiment group) in carrying out the first task, used NetBeans (the control group) to carry out the second task, and vice versa. Each participant carried out each task exactly once individually, either using the TTP Toolkit or using NetBeans. Time to complete the tasks was measured centrally. During the experiments, participants who had technical questions about the use of the tools were answered by the experimenters but they did not receive any indications before or during the experiments as to how to complete the specific tasks. All participants were paid for a fixed duration (five hours) at the rate of £10 per hour even if it took them less to complete the tasks.

Details on each experiment follow. Where advance test runs were conducted with individual participants, their results are included in the analysis of the results of the experiment.

## Experiment 1: comprehension

This experiment was conducted on 14 March 2009 with 10 participants. Participants received one hour training in using NetBeans and one hour training in using the TTP Toolkit in carrying out software comprehension tasks. Only correct results were allowed: once a participant believed they have completed the task, they were asked to write down the answer and show it to the experimenters. If the answer was correct, current time was registered as the completion of the task. Otherwise, the participant's error was indicated to him/her and the participant was asked to proceed in completing the task. The two tasks that participants were asked to carry out are:

- **1st task: understand Java Abstract Window Toolkit**. Participants were provided four of the source code files from package `java.awt` of the standard Java library (JDK 1.6) and the respective javadoc files. Participants were then asked to list four methods in class `Container` that satisfied two specific conditions.

At the conclusion of the first task, participants in the experiment group (using the TTP Toolkit) were moved to the control group (using NetBeans) in carrying out the second task, and vice versa.

- **2nd task: understand Java InputStreams**. Participants were provided four of the source code files form package `java.io` of the standard Java library (JDK 1.6) and the respective javadoc files. Participants were then asked to list four methods in class `BufferedInputStream` that satisfied two specific conditions.

**Results**: Analysis of the results show that, on average, programmers who used the TTP Toolkit required 0.24 of the time it took to complete same task for programmers who used NetBeans and javadoc. The results reported are statistically significant (p-value 0.02).

## Experiment 2: conformance (design verification)

This experiment was conducted on 28 March 2009 with 8 participants. Participants received one hour training in using NetBeans and one hour training in using the TTP Toolkit in carrying out software design verification tasks. Participants were presented with a summary from the description of a design pattern taken from [Gamma et al. 1995] and with an implementation that may or may not conform to the aforementioned description. Users of the TTP Toolkit were also presented with a formal (LePUS3) specification of the design pattern, and were asked to use the TTP Toolkit to verify conformance. At the conclusion of each task, each participant was asked to record whether the verification was successful (i.e., whether the implementation conforms to the specification), to record the time it took him/her to reach the conclusion, and to rate their confidence in the result. The two tasks that participants were asked to carry out are the following:

- **1st task: check conformance to the Composite pattern**. Participants were provided with a summary of the description of the Composite design pattern and source code from package `java.awt` of the standard Java library (JDK 1.6). They were asked whether a named subset of these classes and a named subset of their methods constitute an implementation that conforms to the design pattern. The correct answer was that the implementation (named classes and methods) indeed conforms to the specification.

At the conclusion of the first task, participants in the experiment group (using the TTP Toolkit) were moved to the control group (using NetBeans) in carrying out the second task, and vice versa.

- **2nd task: check conformance to the Decorator pattern**. Participants were provided with a summary of the description of the Decorator design pattern and source code from package `java.io` of the standard Java library (JDK 1.6. They were asked whether a named subset of these classes and a named subset of their methods constitute an implementation that conforms to the design pattern. The correct answer was that the implementation (named classes and methods) does not conform to the specification.

**Results**: Analysis of the results shows that the answers given by the users of the TTP Toolkit were, on average, 2.67 more correct than the answers given by who have used NetBeans (p-value: 0.10). It also shows no significant differences in the time it took to complete the tasks or in the confidence that participants expressed in the correctness of their answers.

## Experiment 3: evolution

This experiment was conducted on 25 April 2009 with 6 participants. Participants received one hour training in using NetBeans and one hour training in using the TTP Toolkit in carrying out software evolution tasks. Participants were provided with a set of source code files and were asked to identify how to make changes to it so that it conforms to specific conditions. At the conclusion of each task, each participant was asked to record the the time it took him/her to complete the task. The two tasks that participants were asked to carry out are:

- **1st task: evolve the InputStream hierarchy**. Participants were provided with a subset of the classes of package `java.io` (from JDK 1.6) and were asked to find the manner by which a specific method with a specific body can be added so as to add a specific behaviour that is in common to two named classes.

At the conclusion of the first task, participants in the experiment group (using the TTP Toolkit) were moved to the control group (using NetBeans) in carrying out the second task, and vice versa.

- **2nd task: evolve the Writers hierarchy**. Participants were provided with a subset of the classes of package `java.io` (from JDK 1.6) and were asked to find the manner by which a specific method with a specific body can be added so as to add a specific behaviour that is in common to two named classes.

**Results**: Analysis of the results reveal that there was no significant discrepancy in the time it took to complete the tasks between the TTP Toolkit and NetBeans users.

## Summary and conclusions

1. Significant improvement was registered in software comprehension tasks, even with very short training time and small implementations. Higher gains are expected with larger programs and experience with the TTP Toolkit as users' proficiency in the tool improves. In addition, a commercial version of the current prototype which overcomes its bugs and current limitations is expected to deliver even greater gains.
2. One hour training in using the TTP Toolkit was not sufficient to shorten the time required to verify conformance to design specifications. However, the TTP Toolkit did increase significantly (more than doubled) the correctness of the verification process.
3. The software evolution experiment has delivered an unexpected result, showing that the TTP Toolkit did not deliver visible gains in carrying out the specified tasks. The reason is most likely because the tasks were too "simple" and the time to complete them was very short (around 5 minutes on average). We hope to conduct another experiment in software evolution during which more complex tasks will be carried out and, we are confident, the TTP Toolkit's advantages will become more evident.